\documentclass[10pt,notitlepage]{revtex4}

\usepackage[latin1]{inputenc}
\usepackage{graphicx}
\usepackage{amsmath,amssymb}
\usepackage{mathrsfs}
\usepackage[percent]{overpic}
\usepackage{mathrsfs}
\usepackage{xcolor}
\usepackage{float}
\usepackage[export]{adjustbox}[2011/08/13]
\usepackage{cancel}
\usepackage[percent]{overpic}

\def\pd{\partial}
\def\cross{\times}

\renewcommand*{\pd}{\partial}
\newcommand*{\fb}{\frac}

\renewcommand{\Delta}{\triangle}

\newcommand{\be}{\begin{equation}}
\newcommand{\ee}{\end{equation}}
\newcommand{\bal}{\begin{align}}
\renewcommand{\Re}{\text{Re}}

\newcommand{\We}{\text{We}}
\newcommand{\Oh}{\text{Oh}}
\newcommand{\tcb}{\textcolor{black}}

\begin{document}

\title{Density contrast matters for drop fragmentation thresholds at low Ohnesorge number}
\author{Florence Marcotte}
\author{Stéphane Zaleski}
\affiliation{Sorbonne Universit\'e, CNRS, Institut Jean Le Rond $\pd$'Alembert, F-75005 Paris, France}

\begin{abstract}
We address numerically the deformation and fragmentation dynamics of a single liquid drop subject to impulsive acceleration by a unidirectional gas stream. The density ratio between the liquid and gaseous phases is varied in the $10-2000$ range and comparisons are made with recent drop breakup experiments. We show for low Ohnesorge numbers that the liquid-gas density contrast significantly modifies the critical Weber number for the transition between bursting and stripping fragmentation regimes, on the one hand, and for drop fragmentation, on the other hand. We suggest a simple theoretical argument to predict the transitional Weber number as a function of the density contrast and show that the stabilising influence of small contrasts can be explained by inertial effects in the nonlinear coupling between drop stretching and centroid acceleration.
\end{abstract}

\maketitle

\textcolor{black}{Drop fragmentation} is the complex phenomenon by which single liquid drops break into child droplets of varying sizes, occurring in a large diversity of natural events (e.g. rain) or industrial processes (e.g. fuel injection in propulsion systems). \textcolor{black}{It is also referred to as \textit{secondary atomisation} in the context of the classical jet atomisation problem where, through the development of successive hydrodynamic instabilities, a high-speed liquid jet destabilises into thin liquid sheets prone to fingering and (primary) drop ejection. The further fragmentation of the resulting drops, if any, depends on the flow conditions and eventually determines the droplets size distribution, a subject of many active debates and a key element toward control of a spray mixing rate \cite{Eggers08,VB09,ling16}.}

The fundamental mechanisms underlying drop fragmentation have been addressed by a vast amount of experimental and numerical studies, which considered the breakup of a single, initially spherical drop immersed in a unidirectional high-speed gas stream. 
 Drop breakup experiments carried out in the last few decades are essentially twofold: a liquid drop is either impulsively accelerated by a shock-wave in a dedicated tube \cite{Hsiang95,Faeth95,chou98,dai01} or released sideways in a high-speed cross-flow \cite{liureitz97,cao07,zhao2010,opfer14}. These experiments have established the existence of a variety of deformation and breakup regimes depending on the flow conditions \cite{Hinze55,ranger69,Gelfand74,krzeczkowski80,pilch87,wierzba90}. Due to practical constraints however, drop breakup experiments have mostly focused on liquid drops whose density $\rho_L$ was large compared to that of the ambient gas $\rho_G$, with density contrasts ${\tt r_d}=\rho_L/\rho_G$ typically larger than 500 (as in the case of a water-air or ethanol-air system at ambient temperature and atmospheric pressure \cite{opfer14}). Nevertheless, small density contrasts (${\tt r_d} \le 10$) are also relevant for combustion sprays due to the high fuel injection pressures involved e.g. in rocket propulsion systems \cite{Heywood88}. Moreover, information such as the thickness of the sheet of inflated drops remains elusive as yet, due to visualisation difficulties. 

Such information, on the other hand, could be easily accessed by means of Direct Numerical Simulations (DNS). Yet numerical studies of drop breakup are limited by the considerable diversity of spatial scales involved as the drop undergoes dramatic stretching and bursting, resulting in huge computational costs, and have long been restricted to small-to-moderate density contrasts with ${\tt r_d}$ typically in the $1-80$ range \cite{zaleski95,han01,kekesi14,yang16,jalaal14}. As a result there is so far little overlap between the flow regimes explored in experiments on the one hand, in simulations on the other hand; and whereas the influence of flow conditions on secondary atomisation has received most attention (especially in terms of the Weber number defined below), this discrepancy arguably contributes to obscure that of the density contrast \cite{aalburg,yang16}. Early simulations  \cite{zaleski95,han01} already suggested that the deformation modes exhibited by the liquid drop at low $\tt r_d$ strongly differ from those observed for similar flow conditions at large ${\tt r_d} \sim 1000$: therefore it appears desirable to bridge the gap between these two parameter regimes and address the sensitivity of the drop breakup dynamics to the density contrast as the latter is varied over a broad range. 

Also of particular interest for industrial applications is to determine the threshold (in terms of flow conditions) beyond which liquid drops become unstable to fragmentation.
In the literature this threshold is characterised by means of two dimensionless parameters, namely the Weber number $\We=\rho_G R_0 U_0^2/\sigma$ quantifying the ratio of inertial to surface tension forces, and the Ohnesorge number quantifying the effet of liquid viscosity ${\Oh}=\mu_L/\sqrt{\sigma \rho_L R_0}$. (Importantly, the latter has been however reported to matter only if exceeding a typical value $\Oh >0.1$.) Here $U_0$ is the velocity difference between the gas stream and the drop, $R_0$ the (undeformed) drop radius, $\mu_G$ the dynamic viscosity of the gas and $\sigma$ the surface tension associated with the gas-liquid interface. Various breakup criteria involving these two parameters are commonly used in non-DNS (Eulerian or Lagrangian) CFD models, where drop breakup is handled with a statistical approach. These numerical criteria either rely on experimental observations (such as the Schmehl model \cite{schmehl} inferred from \cite{pilch87,Hsiang95}) or simple theoretical arguments (such as the Taylor-Analogy Breakup criterion \cite{orourke87}, which uses an analogy between the distorted drop and a spring-mass system \cite{Taylor63}). In a fundamental context, \cite{VB09} addressed the fragmentation of a falling raindrop and suggested a simple inviscid model to account for the critical Weber number characterising the transition from oscillatory, reversible deformation to exponential stretching and breakup, which they found to be ${\We_c} \sim 3$, in excellent agreement with available experiments at large density contrasts.

None of these stability criteria, whether empirical (and based on experiments, hence typically implying ${\tt r_d} > 500$) or theoretical (and based on simplified models), involves the density contrast between the gaseous and liquid phases. Yet, the important morphological discrepancy between deformation regimes observed at low $\tt r_d$ (numerically) and those observed at large $\tt r_d$ (mostly experimentally, although simulations at ${\tt r_d}=1000$ have been very recently reported for Weber numbers well above the critical value {for breakup} \cite{jain15,gaurav2018}) suggest that the effect of density contrast on the fragmentation threshold remains an open question, which it is the aim of the present paper to investigate.


\section{Numerical model}
\label{model}


\subsection{Flow configuration}

Simulations were performed using the flow solver Gerris \cite{popinet03,popinet09} in axisymmetric geometry to model the impulsive acceleration of a spherical liquid drop behind a planar shock-wave. Gerris solves for the two-phase, incompressible Navier-Stokes equations using a finite-volume method with dynamical adaptive mesh refinement, which allows for capturing thin flow structures while mitigating computational costs. The interface is advected using a Volume of Fluid method, and the incompressibility constraint is enforced at each timestep by projecting the velocity field onto the divergence-free manifold using a classical Chorin's algorithm \cite{Chorin67,Chorin68} coupled with a multigrid Poisson solver.

Our computational domain is a cylindrical channel with dimensionless radius $R_{cyl}=5$ and length {$L_{cyl}=15$} in the streamwise direction, where the unit length is the drop initial radius $R_0$ {(the channel length was increased whenever necessary up to $L_{cyl}=25$ to observe the successive deformation stages as the drop travels downstream)}. Along with the prescribed axisymmetry condition on the axis, the boundary conditions are free-slip at the outer radius, free outflow at  the channel outlet, and fixed inflow at the channel inlet, where the prescribed incoming gas velocity $U_0$ sets the velocity unit. The drop center is initially located on the axis at distance $d=3$ from the inlet boundary, and the velocity field is initially set to zero over the entire domain. Refinement of the adaptive quadtree mesh was allowed up to $400$ cells per drop radius. It is important to appreciate that the choice of an axisymmetric model filters out the three-dimensional instability mechanisms eventually leading to droplet atomisation, and that the observed breakup occurs essentially once the thickness of a liquid sheet becomes smaller than the minimal grid size. Prescribing the same maximum level of refinement for all simulations (except for convergence checks) therefore provides a unique criterion for numerical breakup throughout the study.

The numerical problem can be entirely described using four independent dimensionless parameters: in addition to the Weber number $\We$ and the density contrast ${\tt r_d}=\rho_L/\rho_G$ defined in the introduction, we define here the Reynolds number as $\Re=\rho_G U_0 R_0/\mu_G$ and the viscosity contrast ${\tt r_v}=\mu_L/\mu_G$, where $\mu_L$ and $\mu_G$ are respectively the dynamical viscosities of the liquid and gas.


\subsection{Concerning initial conditions}

Dealing with the flow initialisation requires particular care, whether considering our initial velocity field (with $u=0$ over the flow domain and $u=U_0$ on the inlet boundary) or the alternative (and equivalent) setups where the velocity field is initialised as $u=U_0$ (resp. $u=0$) in the liquid drop and $u=0$ (resp. $u=U_0$) in the surrounding gas (with respectively free inflow boundary condition and fixed inlet velocity). It should be emphasized that the numerical model aims at reproducing a physical problem (the impulsive acceleration of a drop traveling across a shock-wave) which as such is a compressible one while discarding the timescale of the acoustic waves. Therefore it necessarily takes a timestep for the numerical solution to adjust and for a (nearly) dipolar velocity field to set in around the drop, which satisfies both the incompressibility constraint and conservation of momentum. As a result however, the relative velocity measured after one timestep between the drop centroid and the gas in the far-field has jumped from the prescribed value $U_0$ by an offset $\delta U$ approximately proportional to $\tt r_d^{-1}$, so that the deviation is particularly pronounced for small density contrasts. 

Importantly, this effect is not a numerical artefact but rather arises from the conservation equations. In the reference frame moving with the shock-wave the drop is initialised with velocity $U_*$ and the gas is at rest. As the shock-wave sweeps over the domain a dipolar flow settles in around the spherical drop, which in turn releases a fraction of its momentum as it accelerates the surrounding gas. The overall momentum is conserved: writing $ {\cal{V}}$ the volume of the drop, and $U_1$ the new relative velocity between the drop and the ambient gas after one timestep, we obtain
\be
\rho_L {\cal{V}} U_* = \rho_L {\cal{V}} U_1 +   \rho_G C_M {\cal{V}} U_1,
\ee
where $C_M=\tfrac12$ is the added-mass coefficient for a spherical drop \cite{magnaudet00}, and we assume the volume of gas entrained by the drop to be equivalent to $ {\cal{V}}$. The new relative velocity $U_1$ is then
\be
U_1= U_* \left(1+ \tfrac12 {\tt r_d^{-1}} \right)^{-1},
\ee
so that for $r_d \gg 1$ the velocity offset $\delta U=U_1-U_*$ becomes at leading order $\delta U \sim - \tfrac12 {\tt r_d}^{-1} {U_1}$.

If we assume that the meaningful quantity is the \textit{effective} velocity difference between the drop centroid and the far-field resulting from the interaction with the shock-wave, the flow parameters used in previous numerical studies of secondary atomisation might be reconsidered with some caution. Specifically to overcome this issue, our simulations were performed using corrected control parameters ${\Re}_*, \We_*$ built on the anticipated value $U_*$, which we determine by requiring the effective velocity difference $U_1$ be kept fixed for all the density contrasts considered, and equal to $U_0$. The timeseries for the centroid velocity shown in section \ref{results} are rescaled accordingly, and all the results presented here therefore correspond to the same, consistent definition of the effective $\We$ and $\Re$ numbers.


\section{Results}
\label{results}


\subsection{Deformation at high density contrasts ; comparison with bag breakup experiments}

The early stages of drop deformation at large $\tt r_d$ are characterised by radial stretching and flattening in the streamwise direction. The drop fate (whether fragmentation or not) then depends on the flow conditions: the restoring effect of surface tension can either overcome the stretching - in which case the drop undergoes oscillatory deformation around a spherical shape while traveling downstream-, or the drop stretches further, inflates and eventually breaks up, displaying a diversity of fragmentation regimes. In the two last decades a vast literature has been dedicated to the experimental investigation of these various deformation modes. As argued in the introduction, most of these experiments were performed in conditions such that the density contrast was typically larger than $500$ and the viscosity contrast $\tt  r_v$  typically in the $100-1000$ range. In this context, the selection of a particular deformation mode was found to be determined by the Weber number \cite{Hinze55,ranger69,Gelfand74,krzeczkowski80,wierzba90,cao07,zhao2010}, with additional influence of the Ohnesorge number whenever the latter exceeds 0.1. At low Ohnesorge numbers (i.e. when viscous effects are weak at the drop scale compared to capillary ones), these studies have shown that the drop dynamics transition from a vibrational mode at the lowest $\We$ numbers, where the accelerated drop undergoes capillary oscillations with or without breakup, to so-called \textit{bag breakup} at $6 \lesssim \We \lesssim 15$, where the drop inflates and bursts like a blown balloon, and \textit{shear breakup} at higher $\We$, where strings of droplets are emitted from the thinning drop edge \cite{pilch87}.

Additional distinctions have been introduced between `simple' bag, `multiple' bag or `bag-and-stamen' modes \cite{pilch87}, but also between `backward-facing' or `forward-facing' bag modes, some of these features being observable only transiently, and therefore difficult to clearly discriminate from each other. {Furthermore, the distinction between `simple' or `multiple' bag deformation can be biased in numerical studies by the neglect of 3D flow features as shown by \cite{jain15}.} In point of fact such a regime diversity may seem unnecessary from a comprehensive point of view, and for the purpose of the present paper we will restrict ourselves to the distinction between three generic regimes, referred to as \textit{oscillatory}, \textit{bursting} or \textit{stripping}, which 
are expected to eventually determine the child droplets size distribution. 
Specifically, the \textit{bursting} regime is characterised by the formation of a well-defined toroidal rim at the edge of the flattening drop and the inflation and rupture of the central liquid sheet, whatever the curvature of the sheet and the location of its first rupture point (here bursting regime therefore includes any type of bag breakup, whether `simple', `multiple', `backward-' or `forward-facing bag'). The \textit{stripping} regime on the other hand corresponds to the progressive mass loss associated with small droplets being emitted away from the thinning edge (this regime has been also referred to as `sheet thinning' or `shear breakup' in the literature). Finally, oscillatory regime here denotes the absence of any fragmentation process. \textcolor{black}{Note that this attempted classification builds on a phenomenological description of the deformation and breakup dynamics and by no means on an assumed distinction between the physical mechanisms underlying fragmentation.} \textcolor{black}{The mechanisms for sheet piercing, whether associated with aerodynamic effects such as the development of a flapping instability, an acceleration-driven hydrodynamic instability, or even the presence of impurities in the drop liquid, are poorly understood and remain an important subject of investigation.}

\begin{figure}[h]
\begin{center}
\includegraphics[width=0.5 \textwidth,trim=0 0 270 0,clip=true]{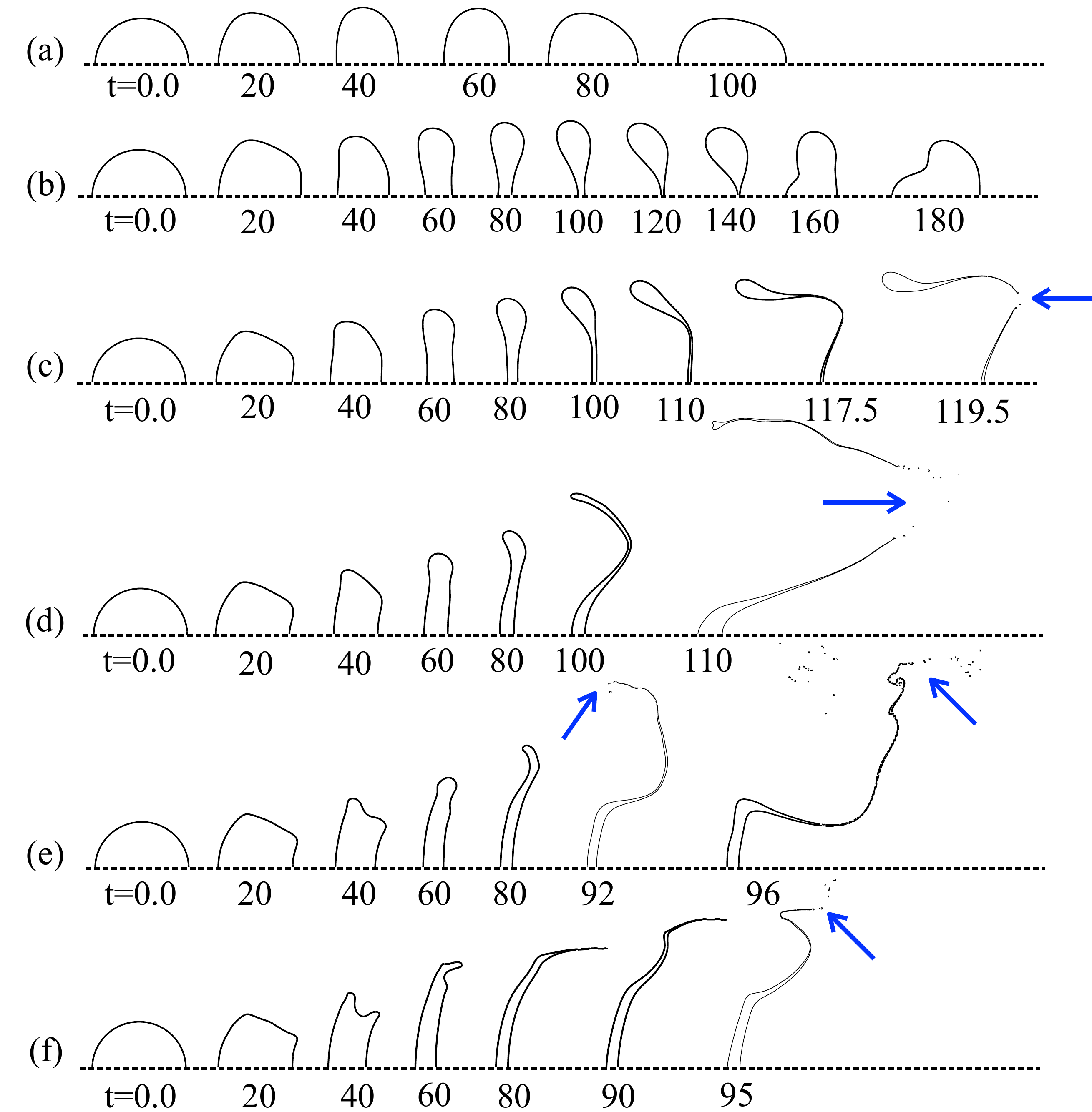}\\
\vspace{1ex}
\includegraphics[width=0.5\textwidth,trim=0 500 0 80,clip=true]{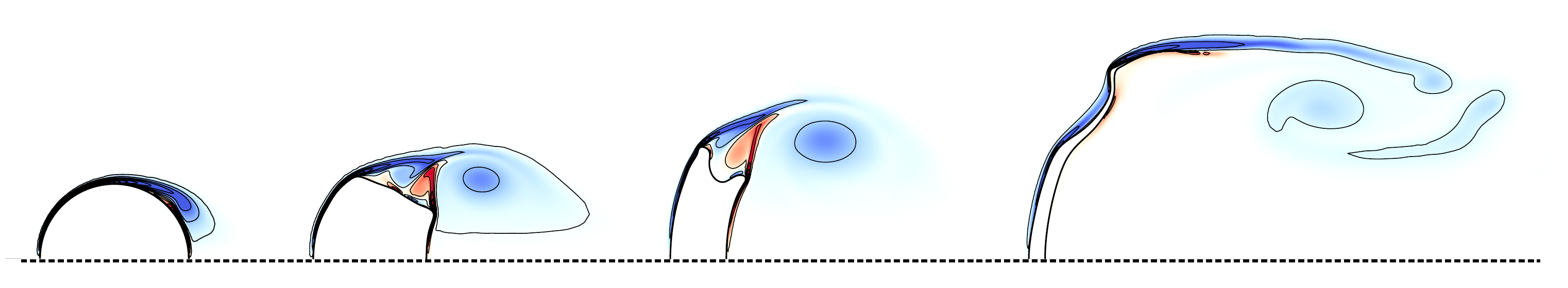}
\caption{\small{Drop deformation for increasing Weber number \textit{(a-f)}: $\We=$  $2.5; 5.65; 7.5; 12; 20; 40$. {Note that the simulations are axisymmetric and the full drop volume is obtained by revolution of the sketched drop section around the dashed (symmetry) axis.} The density contrast ${\tt r_d}=1110$, Reynolds number $\Re=1090$ and viscosity contrast ${\tt r_v} =90.9$ are fixed ; here the gas stream blows from the left side.  Dimensionless times {$\tilde{t}=t\, U_0/R_0$} are indicated on each snapshot. \textit{Bottom line:} snapshots of the vorticity field in the gas (isolines) for case \textit{(f)}, shown here at times $\tilde{t}=0; 20; 40; 90$. Blue arrows highlight the region where breakup first occurs. \tcb{Note that even though our axisymmetric simulations cannot truly reproduce the 3D breakup mechanisms with the expansion of circular holes in the bursting drop sheets, the very occurence of a breakup event {(whether bursting, or stripping at higher We)} is not a numerical artefact and is also observed in experiments}.}}
\label{fig:We1}
\end{center}
\end{figure}

Figure \ref{fig:We1} illustrates the drop typical morphological evolution corresponding to these three regimes, as can be observed successively by increasing $\We$ at fixed \{${\tt r_d}=1110$, ${\tt r_v}=90.9$, $\Re=1090$\} in our simulations, where the contrasts between fluid properties ${\tt r_d}$ and ${\tt r_v}$ are typically close to that of an air-water system at $293~{\rm K}$. The two examples of oscillatory deformation shown in figure \ref{fig:We1}(a) and (b) respectively correspond to the $\We=2.5$ and $\We=5.65$ cases: the drop flattens and a toroidal rim forms at the edge. Here the case $\We=5.65$ (b) transiently appears to be on the verge of break-up until surface tension effects induce the rim to drain back into the drop core region. The two examples of bursting behaviour in figure \ref{fig:We1}(c) and (d) respectively correspond to what is classically referred to as bag (or backward-facing bag) mode and multiple bag (or even bag-and-stamen) mode, here for $\We=7.5$ and $\We=12$. Both evolutions are characterised by the formation of a well-defined rim at the edge of the flattening drop and the extreme thinning of the liquid sheet in the interior, eventually causing the inflated drop to burst. Bursting occurs as the bag thickness locally becomes smaller than the minimal grid size, and is typically initiated in the vicinity of the rim neck. It has been argued that the selection of a simple or multiple bag deformation mode could be associated to different most unstable wavelengths with respect to Rayleigh-Taylor instability \cite{theofanous04}. In point of fact, whether `simple' or `multiple' bag is seen here to result from an inertia competition between the outer, heavy region forming the rim and the inner sheet at the drop center, resulting in one region being accelerated faster than the other. The size and retractation speed of the rim \cite{Taylor59c,Culick60} being controlled by the surface tension, it is naturally expected that larger $\We$ (hence smaller, lighter rims) will result in the rim neck moving outward and the inner region becoming heavier than the rim. 

Finally, two stripping-mode examples are shown in figure  \ref{fig:We1}(e) and (f) respectively for $\We=20$ and $\We=40$: here the rim is hardly (if at all) visible, and the early stages of drop deformation exhibit a typical `axe-like' shape (as in the snapshots shown at $\tilde{t} \sim 40$). This particular feature progressively becomes more acute for larger $\We$ as the smoothing operated by the weaker surface tension cannot counteract the erosion of the drop by its own vortical wake (as illustrated by the vorticity fields in figure \ref{fig:We1} for $\{\We=40 \, ; \, \Re=1100 \, ; \, {\tt r_d}=1110 \, ; \, {\tt r_v}=90.9\}$ ; bottom line). As the drop further flattens the indentation carved by the vortical wake tightens and shuts down, until its remnant becomes the noticeable irregularity of the drop edge shown at $\tilde{t}=60$ in figure \ref{fig:We1}\textit{(f)}. Fragmentation eventually occurs as small droplets are stripped away repeatedly from the thinning edge of the dramatically stretching drop, which behaves like a downstream-facing bag. The transition from bursting to stripping mode appears to be a continuous one and corresponds to the disappearance of the rim.\\

\begin{figure}[h]
\begin{center}
\centerline{
\raisebox{44mm}{\rotatebox{90}{\large{$R/R_0$}}}
\includegraphics[width=0.55\textwidth,trim=40 0 100 0,clip=true]{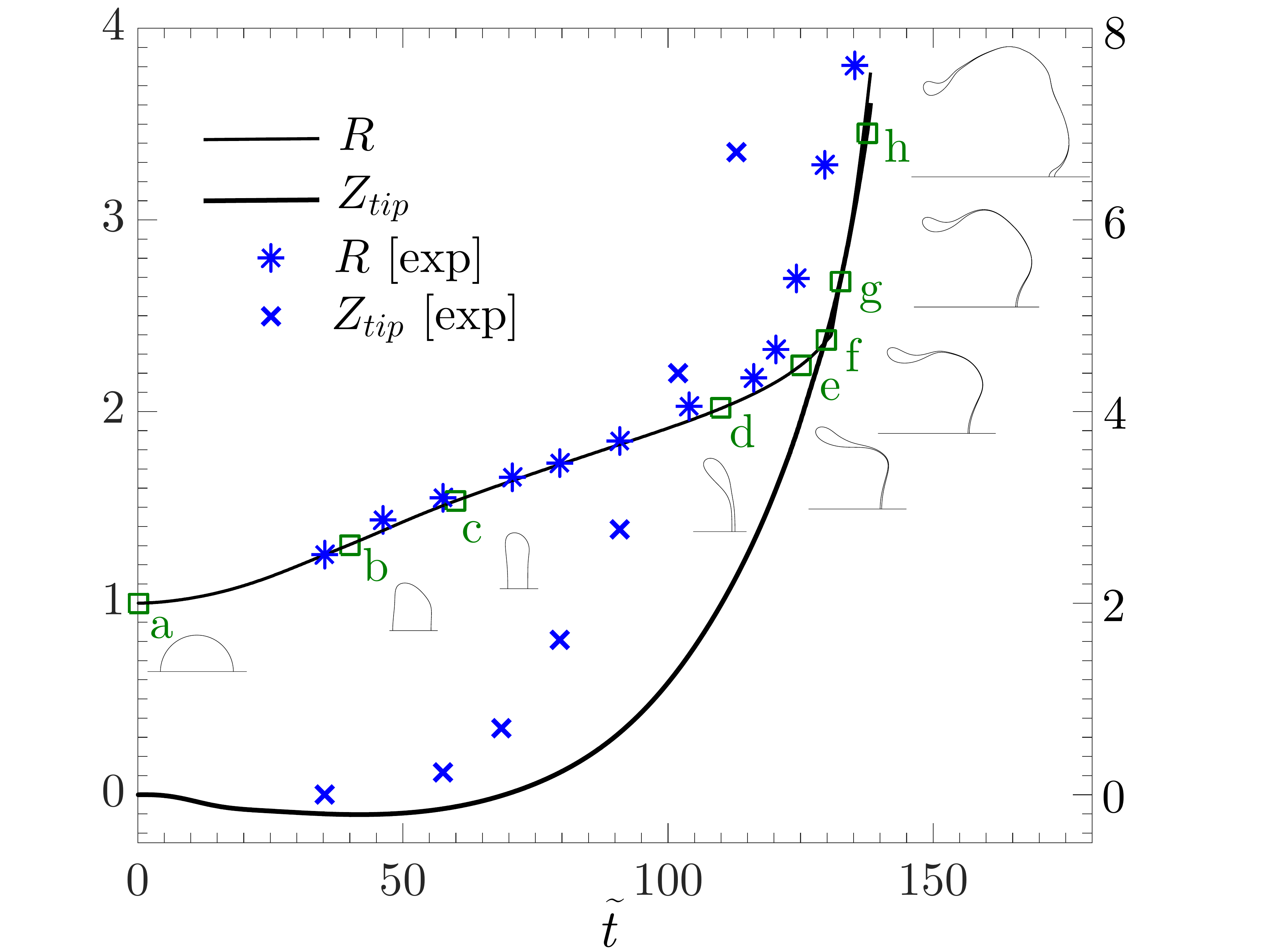}
\raisebox{52mm}{\rotatebox{-90}{\large{$Z_{tip}/R_0$}}}
\hspace{4ex}
\includegraphics[width=0.35\textwidth]{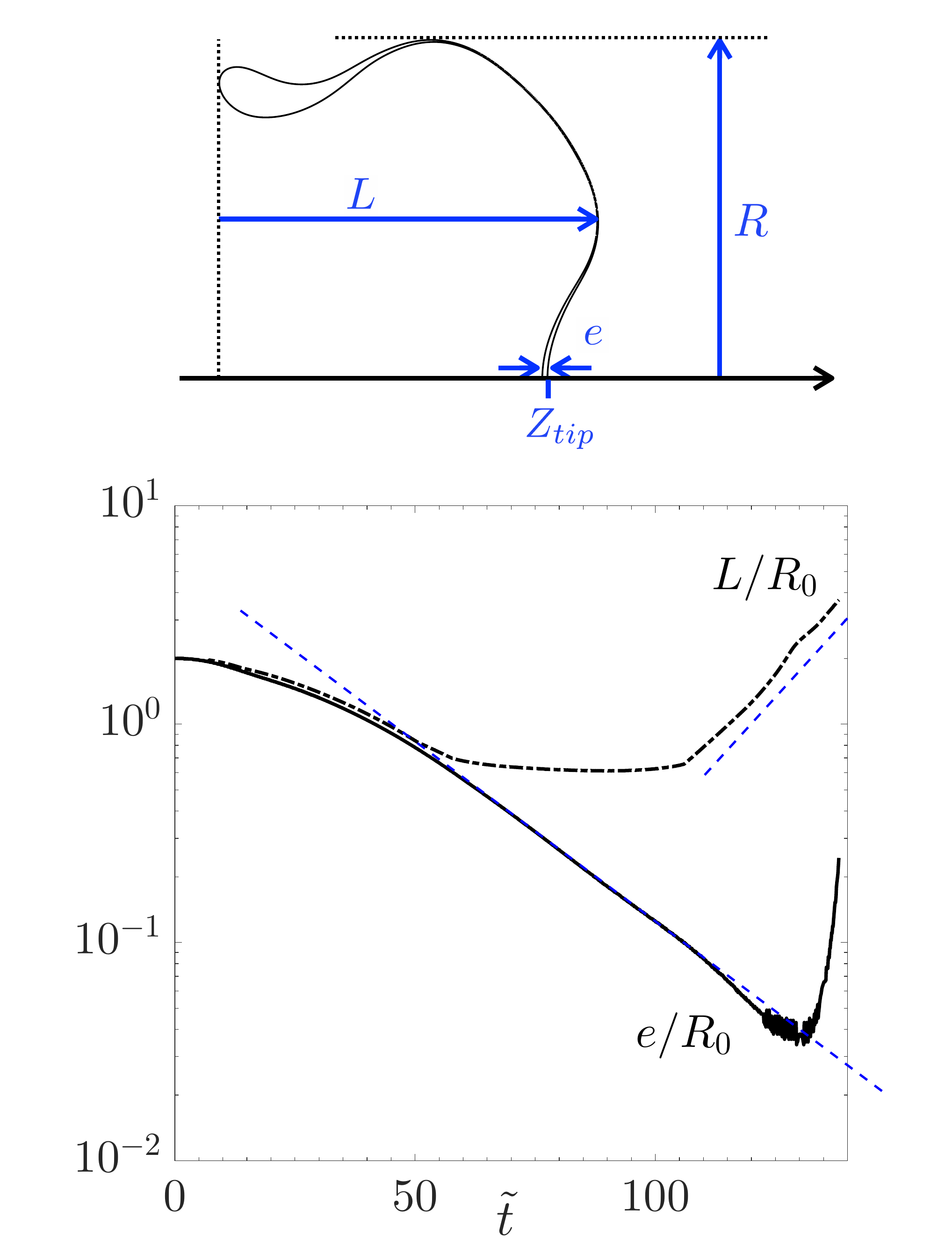}
}
\caption{\small{\textit{Left:} Displacement $Z_{tip}$ of the drop tip along the $z$ axis (thick solid line) and drop radius $R$ (thin solid line) as a function of the dimensionless time {$\tilde{t}=t \, U_0/R_0$}, for $\Re=1484$, $\We=7.55$, ${\tt r_d}=1110$ and ${\tt r_v}=909$. The plotted quantities are defined in the schematic \textit{(top, right)}. As illustrated by the snapshots (given at the times signaled by the letters `a' to `h'),  the deformation regime corresponds here to (single) bag bursting. The conditions reproduce experimental ones using ethylene glycol drops in \cite{opfer14} (case \textit{e-3}): data extracted from their experimental results (\cite{opfer14}, figure 4) are superimposed for reference (blue stars: drop radius, blue crosses: displacement of the tip). The data are recast in terms of our dimensionless variables using the values given in (\cite{opfer14}, table 2) for the length and velocity units: $R_0= 1.35\text{mm}$ and $U_0=15\text{m/s}$. Note that in \cite{opfer14}, the `initial' drop radii (first experimental points) in their figure 4 do not match those indicated in their table 2: we have assumed that the first deformation stage (during which the drop travels through the boundary layer toward the center of the channel) was not recorded in their timeseries. Therefore we have arbitrarily shifted the origin of their times to the time at which the drop radius was numerically observed to match that of the first experimental point. The same timeshift was then applied for the (experimental) evolution of the tip displacement. \textit{Right, bottom:} Drop thickness on the axis $e$ (solid line) and bag length $L$ (dotted-dashed line) versus dimensionless time. The thin, dashed lines emphasize stages of exponential growth or decay.
}}
\label{fig:B24}
\end{center}
\end{figure}

Figure \ref{fig:B24}\textit{(left)} shows the displacement of the drop tip along the streamwise direction (thick solid line) and the drop radius (thin solid line) as a function of the dimensionless time {$\tilde{t}=t\, U_0/R_0$}. Here the particular fluid properties ($\tt r_d, r_v$) and flow conditions ($\Re,\We$) are chosen so as to reproduce the experimental conditions reported in \cite{opfer14} (see their Figure 4, case e-3). Although some differences arise from the comparison between their experimental results and our numerics (for instance the drop tip position slightly displaces upstream during the first flattening stage, and the slope of the drop radius evolution is close to zero in the numerical results at $\tilde{t}=0^+$; both features are missing in the experimental curves), these discrepancies can be traced back to the difference in flow initialisation. Indeed, in the experiments the drop is introduced sideways into the channel and has to travel across the boundary layer where it already feels (differential) acceleration by the gas stream before reaching the region of uniform, maximal flow intensity. We therefore expect the transient dynamics to differ from what is monitored in our simulations. Nevertheless, and despite the uncertainties related to initialisation, we find the evolution of the drop radius (dotted line) to agree well with experimental measurements (denoted by the blue stars ; the method used to superimpose the experimental data from \cite{opfer14} with our numerics is carefully detailed in the caption of figure \ref{fig:B24}). Both timeseries display a clear transition between a first deformation stage, in which the maximal drop radius grows (almost) linearly between $\tilde{t} \sim 50$ and $\tilde{t} \sim 110$, and a later stage of exponential growth, although the latter is initiated slightly earlier in the experiments ($\tilde{t} \sim 120$) than in the numerics ($\tilde{t} \sim 130$, corresponding to snapshot `e' in \ref{fig:B24}). The snapshots (`f' to `h') show that this later stage corresponds to the catastrophic inflation of the bag, and the maximal radius is no longer that of the outer rim. \textcolor{black}{(Closer observation of the drop radius timeseries and its successive inflexion points during the first deformation stage (up to $\tilde{t} \sim 110$) actually points towards a behaviour similar to that of a half-oscillation (with dimensionless period $\sim 200$), interrupted at later times by a catastrophic inflation stage. This point will be further discussed later on.)}


Numerical and experimental measurements of the tip displacement are more difficult to compare. Although the trends are qualitatively similar, the acceleration of the drop tip seems to be achieved earlier in the experiments - but here again it should be emphasized that in this case the drop has been already progressively accelerated during its travel toward the center of the channel, making the comparison uncertain. An important outcome of our numerics is that they provide for the first time a direct numerical estimate of the drop thickness (solid line in figure \ref{fig:B24}\textit{(bottom, right)}), here measured on the symmetry axis. From $\tilde{t} \sim 60$ (which approximately corresponds to the formation of the rim (snapshot `c') up to $\tilde{t} \sim 110$ (snapshot `d'), the time evolution of the thickness on the axis (which for this time window is also the minimal thickness) is well represented by an exponential decay (dashed line). The film thickening observed at later times corresponds to a stage where the minimal film thickness is no longer achieved on the axis. Finally, the later stage of drop deformation is characterised by an exponential growth of the bag length $L$ (figure \ref{fig:B24}\textit{(bottom, right}, dotted-dashed line), starting around $\tilde{t} \sim 110$ (snapshot `d'). \textcolor{black}{As seen from figure \ref{fig:B24}\textit{(left)}, this time coincides with the end of the first stage in the evolution of the drop radius, as catastrophic bag inflation takes over the `half-oscillatory' behaviour. Breakup occurs (numerically) as the (axisymmetric) sheet minimal thickness reaches the smallest grid size: assuming an initial drop radius of $R_0= 1.35\text{mm}$ as in \cite{opfer14}, this critical thickness would correspond here to $3.2{\mu}m$ (note that in \cite{opfer14} bursting is observed experimentally for a thickness of $\sim1{\mu}m$).}

\begin{figure}[h]
\begin{center}
\includegraphics[width=0.6\textwidth,trim=0 100 0 0,clip=true]{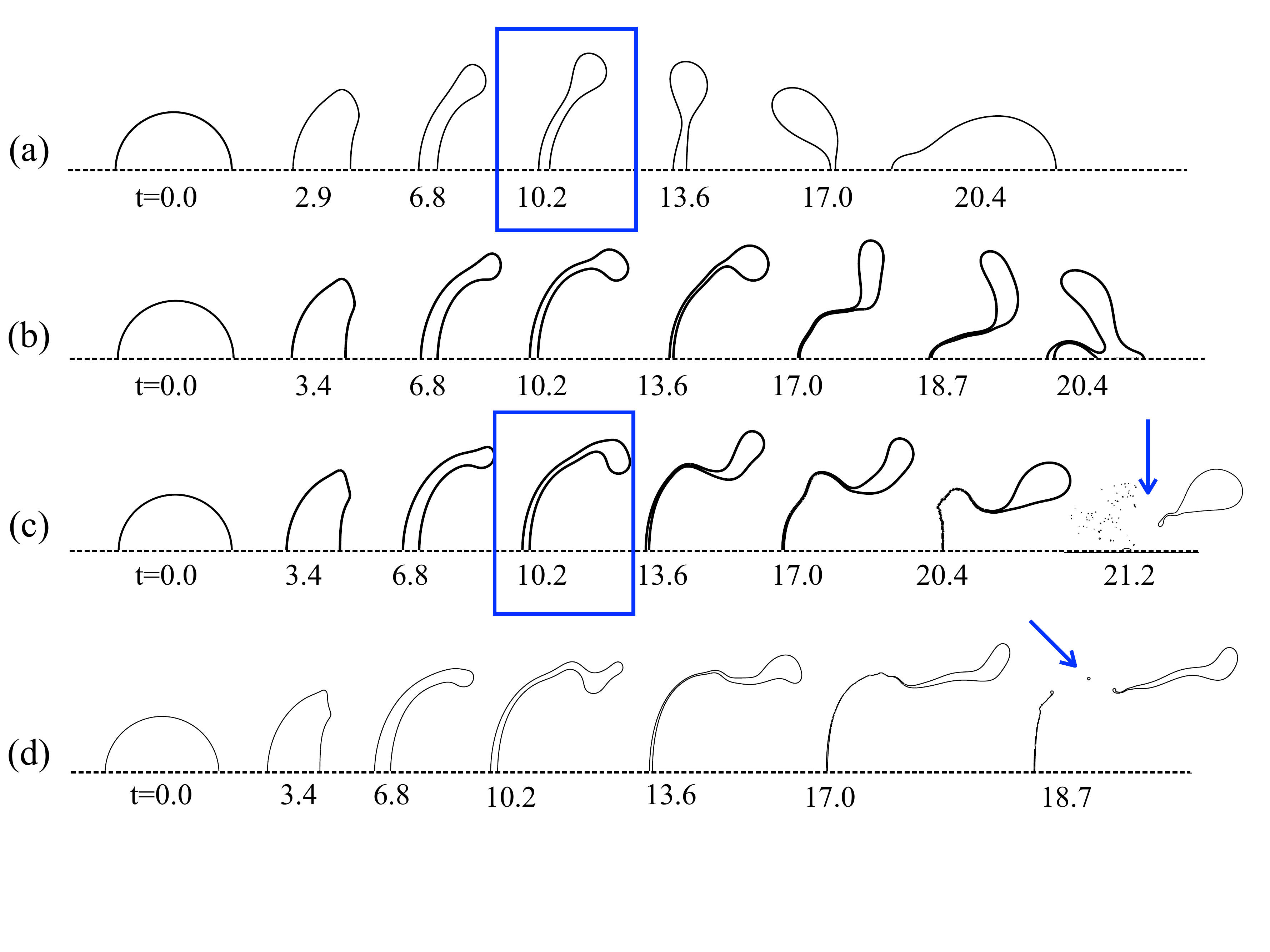}
\includegraphics[width=0.39\textwidth]{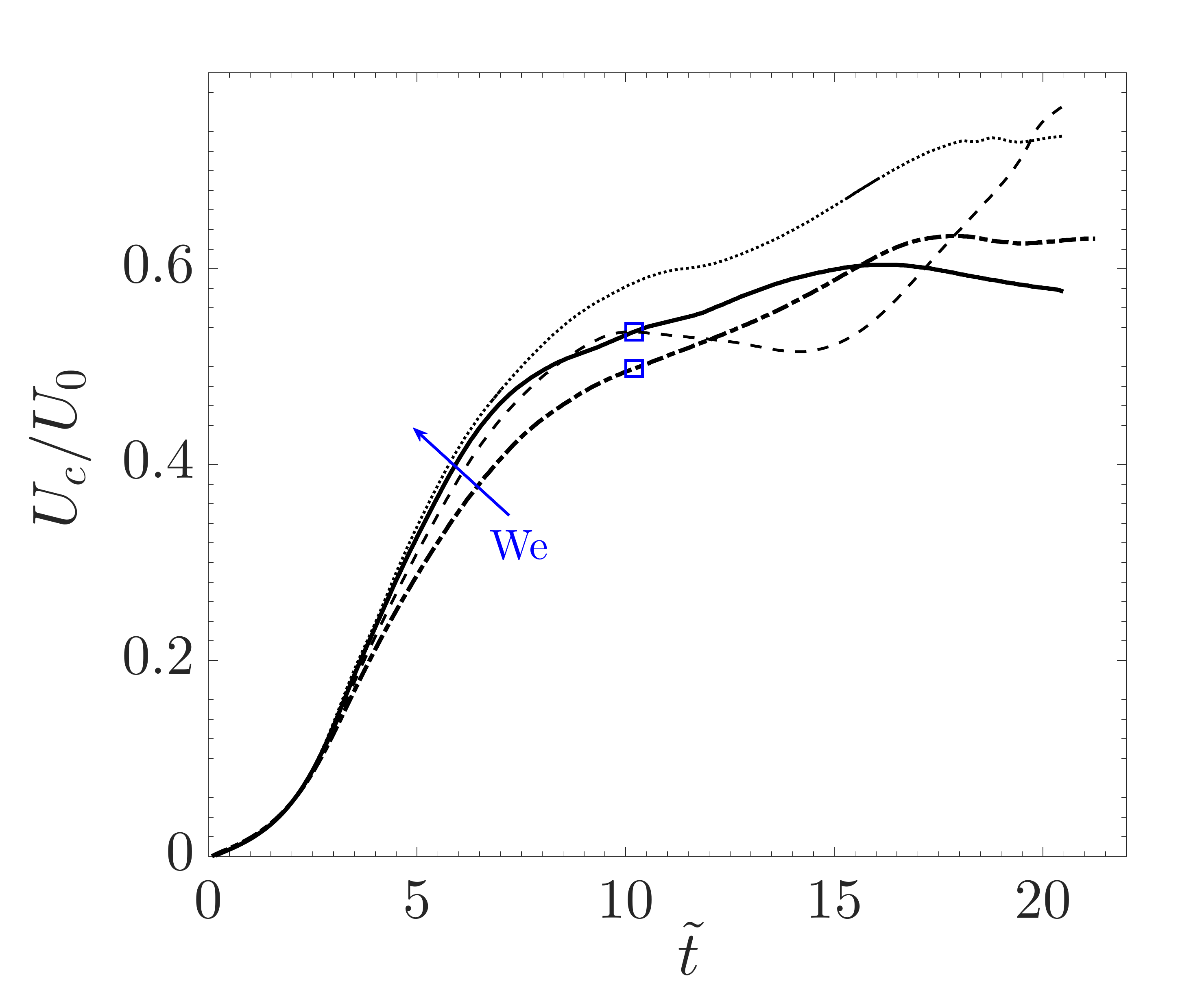}
\caption{\small{\textit{Left:} Drop deformation for increasing Weber number \textit{(a-d)}: $\We=$  $7.5; 12; 20; 40$. The density contrast ${\tt r_d}=10$, Reynolds number $\Re=1090$ and viscosity contrast ${\tt r_v} =90.9$ are fixed. Dimensionless times {$\tilde{t}=t\, U_0/R_0$} are indicated on each snapshot. \textit{Right:} Centroid velocity as a function of the dimensionless time {$\tilde{t}$}, for $\We=~7.5$ (dashed-dotted line), $\We=~12$ (dashed line), $\We=~20$ (dotted line) and $\We=~40$ (solid line). For reference empty squares mark the time of the corresponding framed snapshots.
}}
\label{fig:We2}
\end{center}
\end{figure}


\subsection{From large to small density contrasts ; from bursting to stripping}

Importantly, the evolution of drop deformation for increasing Weber numbers at small density contrast seems to significantly differ from the large density contrast case, as already observed from early drop breakup simulations \cite{zaleski95,han01}. Figure \ref{fig:We2}\textit{(left)} illustrates this evolution for ${\tt r_d}=10$ and $\We$ in the $7.5-40$ range, while the Reynolds number and viscosity contrast are chosen as in figure \ref{fig:We1}. These results are in excellent agreement with previous numerical results of \cite{zaleski95,han01,gaurav2018}. Three observations are readily made. Firstly, we find the transition from oscillatory to fragmentation regime to occur at significantly higher $\We$ compared to the large-density series in figure \ref{fig:We1}. Secondly, for all the $\We$ numbers presented here the drop rapidly bends in the downstream direction and develops (at least transiently) into a downstream-facing bag, which - unlike that of figure \ref{fig:We1}(e) and (f) at high density contrast - exhibits a well-defined rim. No stripping regime could be observed up to $\We=40$, \textcolor{black}{where the inflated drop deforms into a typical jellyfish shape.}   
Thirdly, although a large (and obviously heavy) rim is observed also at large $\We$, the thin (and obviously lighter) inner region seems to lag behind the rim until it either bursts (at high $\We$) or catches up under effect of surface tension at lower $\We$, so that the competition of inertia between the rim and the inner region discussed for the large density contrast case may not seem to apply here. In fact, this apparent contradiction is removed by considering the instantaneous centroid velocity $U_C$ of the drop on figure \ref{fig:We2}\textit{(right)}, \textcolor{black}{where $U_C=1/{\cal V}\int_{\cal V} u d\cal V$. (Note that $U_C$ is expected to differ from the velocity of the rim or that of the drop apex but is considered here for simplification.)} By the time $\tilde{t}=10$ the centroid velocity has already {temporary} reached a (near to) saturation \textcolor{black}{plateau} and the drop does not feel the acceleration {at this stage}. {(The mechanism for temporary stabilisation of the centroid velocity lies in the nonlinear coupling between drop stretching and acceleration, as will be transparent from the simple model developed and discussed later on in section \ref{sec:threshold}.)}

\begin{figure}[h]
\begin{center}
\includegraphics[width=0.65\textwidth,rotate=270]{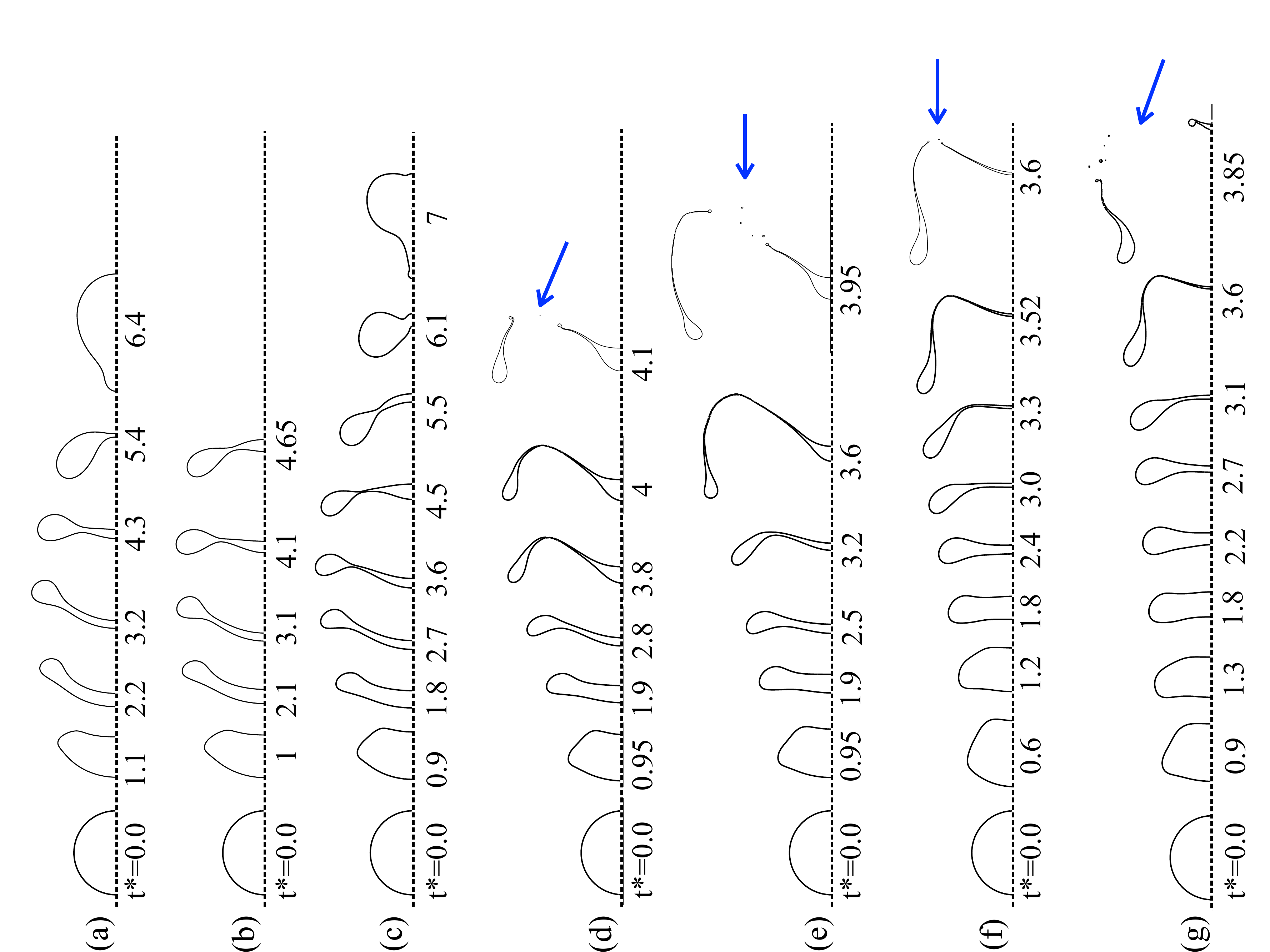}
\caption{\small{Drop deformation for increasing density contrast \textit{(a-g)}: $\tt r_d=$  $10; 20; 40; 111; 250; 1111; 2000$. The Weber number $\We=7.5$, Reynolds number $\Re=1090$ and viscosity contrast ${\tt r_v} =90.9$ are fixed. Dimensionless times are indicated for each snapshot in terms of the reduced time $t^*={\tt r_d}^{-1/2} \tilde{t}$.}}
\label{fig:We7d5}
\end{center}
\end{figure}

Finally, the transition between deformations regimes observed at low and high density contrasts (respectively in early numerical studies 
and experimental ones
 \textcolor{black}{; see for example the references given in Introduction)} is illustrated by the snapshots series in figure \ref{fig:We7d5}, where the Weber number was kept fixed at a small value $\We=7.5$ and the density contrast varied for the first time in the $10-2000$ range: the downstream bending of the flattened drop at early times progressively disappears as the density contrast increases, and is no longer observable from ${\tt r_d} \sim 100-250$. The times corresponding to the various snapshots in figure \ref{fig:We7d5} are expressed in terms of the reduced (dimensionless) time $t^*={\tt r_d}^{-1/2} \tilde{t}$. 
 
\begin{figure}[h]
\begin{center}
\includegraphics[width=0.5\textwidth,trim=0 20 0 0,clip=true]{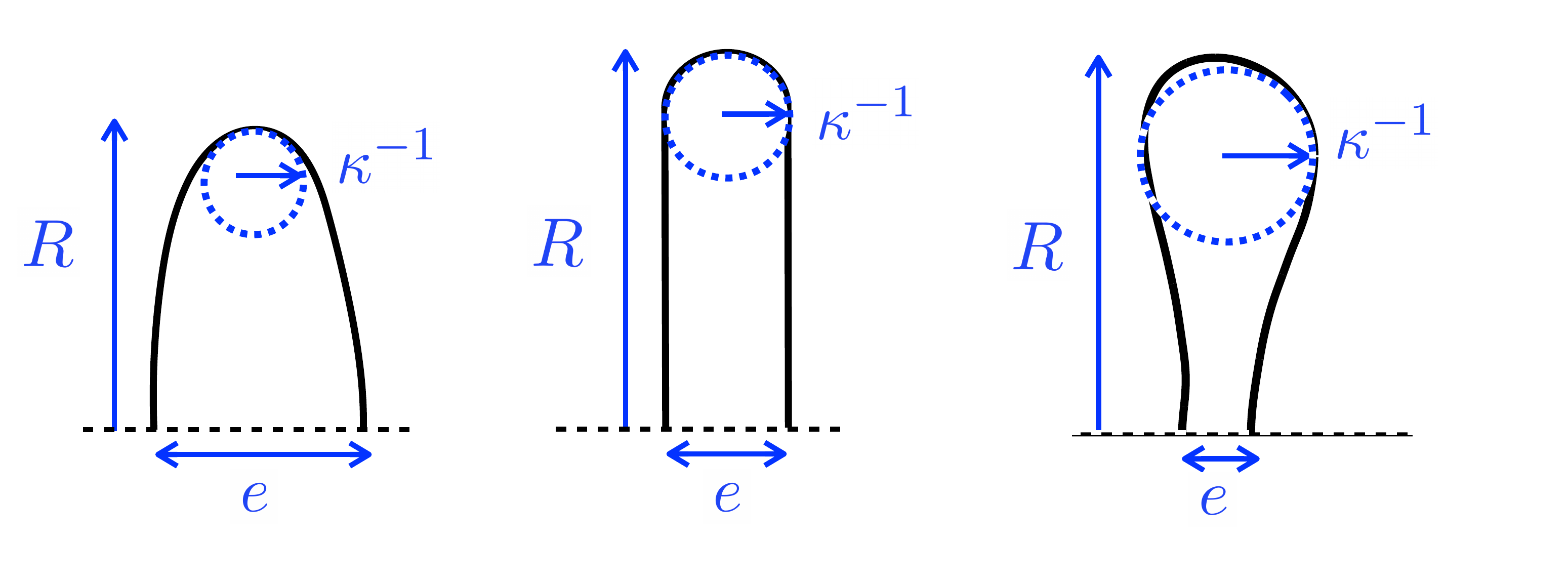}
\caption{{\small{Schematic drawing defining the variables used for the derivation of (\ref{eq3})-(\ref{We2}), in the marginal (idealized) case where the rim is about to develop: $\kappa^{-1} \sim e/2$ \textit{(middle)}, before \textit{(left)} and after \textit{(right)} the formation of a well-defined rim.}}}
\label{fig:figadd}
\end{center}
\end{figure}
 
{A particularly remarkable} feature in the drop deformation dynamics is the reduced time $t^*_{\cross}$ at which the rim starts to develop, which in first approximation appears to be almost invariant with the density contrast (here $t^*_{\cross} \sim 1.8$ was found for $\We = 7.5$). This observation can be simply explained by considering the force balance: discarding viscous effects, a rim forms at the edge of the stretching drop when the restoring effect of surface tension overcomes that of inertia, i.e. when
\be
\label{eq3}
\rho_L \dot{R}^2 \sim \sigma \kappa,
\ee
{where $\kappa$ is the curvature of the rim radius as defined on the schematic drawing in figure \ref{fig:figadd}.} At this stage, the rim radius is equivalent to the drop half-thickness $e/2$, so that the previous condition rewrites as
\be
\label{We2}
 \fb{\rho_L e \dot{R}^2 }{2\sigma}  \sim  1,
\ee
meaning that the Weber number built on the stretching velocity and the drop half-thickness becomes close to unity. Continuity of stagnation pressure across the interface implies that the stretching velocity is related at short times to the gas velocity through $\dot{R} = U_0 \sqrt{\rho_G}/(\sqrt{\rho_L} + \sqrt{\rho_G})$ - referred to as Dimotakis velocity \cite{dimotakis86} -, reducing to  $\dot{R}= U_0 {\tt r_d}^{-\tfrac12}$ for $\rho_G \ll \rho_L$. Using mass conservation in the flattening drop, condition (\ref{We2}) thus becomes
\be
1  \sim \We \, \fb{e}{2 R_0}  \sim \We \, \frac23 \fb{R_0^2}{R^2},
\ee
and finally
\be
1 \sim \We \, \frac23 (1+ t^*_{\cross})^{-2},
\ee
yielding $t^*_{\cross} \sim 1.2$ for $\We =7.5$. Note that this estimate builds on the assumption that $\rho_G \ll \rho_L$ and therefore fails as $\rho_G$ and $\rho_L$ become comparable.

The transition from bursting to stripping regime corresponds to the progressive disappearance of the rim. It is associated with \textcolor{black}{the displacement of the most-thinning region from the interior of the sheet (in the vicinity of the rim neck in bursting regime) toward the outer edge (in stripping regime).} Since the argument above provides a typical timescale for the formation of the rim, it is therefore natural to compare $t^*_{\cross}$ with the typical timescale for viscous entrainment in order to predict the transition between the two regimes. As in \cite{kekesi14}, the {dimensional} `shear' breakup timescale $t_S \sim e/U_L$ (with $U_L$ the velocity close to the interface inside the liquid drop) can be estimated using the continuity of viscous stress across the interface:
\be
\mu_G \fb{U_0}{\delta} \sim \mu_L \fb{U_L}{e},
\ee
where $\delta \sim \sqrt{\mu_G R_0/ \rho_G U_0}$ is the typical thickness of the laminar boundary layer in the gas \cite{blasius08}, and for simplicity $e$ is assumed to provide a relevant lengthscale for the laminar boundary layer in the liquid. The reduced, dimensionless shear breakup timescale is therefore approximated by the relationship
\be
t^*_S = {\tt r_d}^{-\tfrac12}  {{\fb{U_0}{R_0}}} t_S \,  \sim \,  {\tt r_d}^{-\tfrac12} {\tt r_v} \sqrt{\fb{1}{\Re}}.
\label{tstarS}
\ee

In \cite{kekesi14}, the argument suggested to predict the transition from bursting to stripping regime is equivalent to comparing the typical reduced (dimensionless) time for shear breakup $t^*_S$ with unity, or comparing its dimensional counterpart with the typical stretching timescale, built on Dimotakis velocity and the drop undeformed radius. However, this approach was motivated by a numerical study performed at constant $\We$ and does not take the surface tension into account. 
Here we compare instead the typical reduced timescale on which shear stresses operate, $t^*_S$, with the typical reduced timescale for rim formation $t^*_{\cross}$: even though the latter is (approximately) independent on the density contrast, the former is not, so that the $\We$ number characterising the transition between bursting and stripping regimes is found to scale like
\be
\We \, \sim \, \frac32 \left(1+ \lambda \, {\tt r_d}^{-\tfrac12} {\tt r_v} \sqrt{\fb{1}{\Re}} \right)^2,
\label{transitionBS}
\ee
{where $\lambda$ is the prefactor inherited from (\ref{tstarS}).} The predicted transition threshold does indeed decrease with increasing density contrasts, consistently with our observations{; moreover, this threshold asymptotically tends toward a constant value at large $\tt r_d$, in agreement with the phase diagram presented later on in figure \ref{fig:diagphi}. We also note that \cite{Hsiang95} suggest a prediction $\We \sim \sqrt{Re}$ for a somewhat related transition from ``bowl-'' to ``dome''-shape regime. This prediction however builds on different theoretical considerations to which the authors do not subscribe; in particular, the authors  of \cite{Hsiang95} use the undeformed drop radius $R_0$ throughout their analysis, whereas we model the variation of the radius $R(t)$ in time.}

\subsection{Fragmentation threshold} \label{sec:threshold}

As argued in the Introduction, determining the effect of the liquid-gas density contrast on the fragmentation threshold at low Ohnesorge number has important implications e.g. for CFD modeling purposes, but has not been investigated so far due to the missing overlap between numerically accessible parameters and explored experimental conditions. Taking advantage of Gerris's capability of accurately handling large $\tt r_d$, the influence of the density contrast on the critical $\We$ number for fragmentation was characterised for $\tt r_d$ varying by nearly 3 orders of magnitude.

\begin{figure}[h]
\begin{center}
\includegraphics[width=0.7\textwidth,trim=40 0 0 0,clip=true]{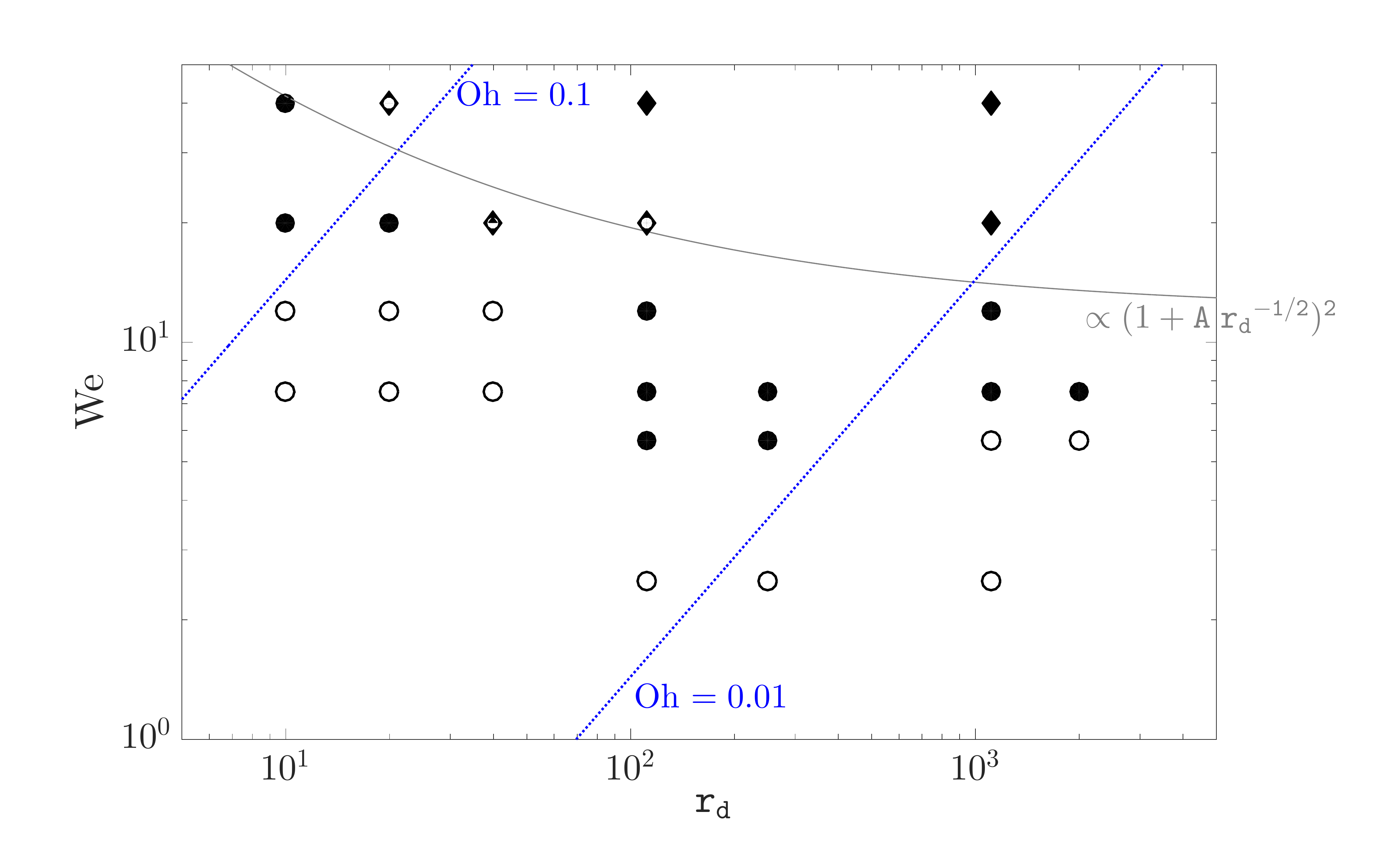}
\caption{\small{Phase diagram determined numerically for constant $\Re=1090$ and ${\tt r_v}= 90.9$, in the ${\tt r_d}-\We$ plane: full symbols denote fragmentation and empty symbols denote oscillatory deformation regime (without fragmentation). Among the full symbols, discs correspond to bursting regime whereas diamonds correspond to stripping regime. \textcolor{black}{Mingled diamonds/discs symbols correspond to transitional regime.} The corresponding Ohnesorge numbers ($\Oh={\tt r_v}/\Re\sqrt{\We/{\tt r_d}}$) are small and lie within the $\Oh=0.004-0.17$ range (the dotted isolines $\Oh=0.1$ and $\Oh=0.01$ are given for reference). {Additionally, the thin, solid line provides a comparison with our prediction (\ref{transitionBS}) for the transitional $\We$ between bursting and stripping regime: the prefactors being unknown, we have arbitrary prescribed ${\tt A}={\tt r_v}/\sqrt{\Re}$, a choice equivalent here to $\lambda=1$ in (\ref{transitionBS}).}
}}
\label{fig:diagphi}
\end{center}
\end{figure}

Figure \ref{fig:diagphi} presents the phase diagram obtained for fixed gas Reynolds number $\Re=1090$ and viscosity ratio ${\tt r_v}=90.9$ in the ${\tt r_d}-\We$ plane, where empty symbols denote oscillatory mode (no breakup) and full symbols denote atomisation. \textcolor{black}{Among the latter, discs correspond to bursting and losanges to stripping regime. Mingled symbols on the other hand correspond to a transitional regime where the drop first undergoes dramatic edge-thinning (thus similar to stripping regime dynamics) before a rim starts to form on the outer edge, and breakup first occurs at a thinner region in the sheet interior.} \textcolor{black}{Our most important finding here} is that the fragmentation threshold drops by one order of magnitude as the density contrast increases from ${\tt r_d}=10$ to ${\tt r_d} =250$. In the range ${\tt r_d} = 250-1000$ the critical value lies between $2.5$ and $5.65$, consistently with the prediction of $\We_c \sim 3$ formulated by \cite{VB09} in the case of a air-water (rain) system (${\tt r_d} \sim 1000$). At larger contrast (${\tt r_d}=1000-2000$) the fragmentation threshold is found to rise again, although by a less significant amount (a factor two).

\begin{figure}[h]
\begin{center}
\includegraphics[width=0.8\textwidth]{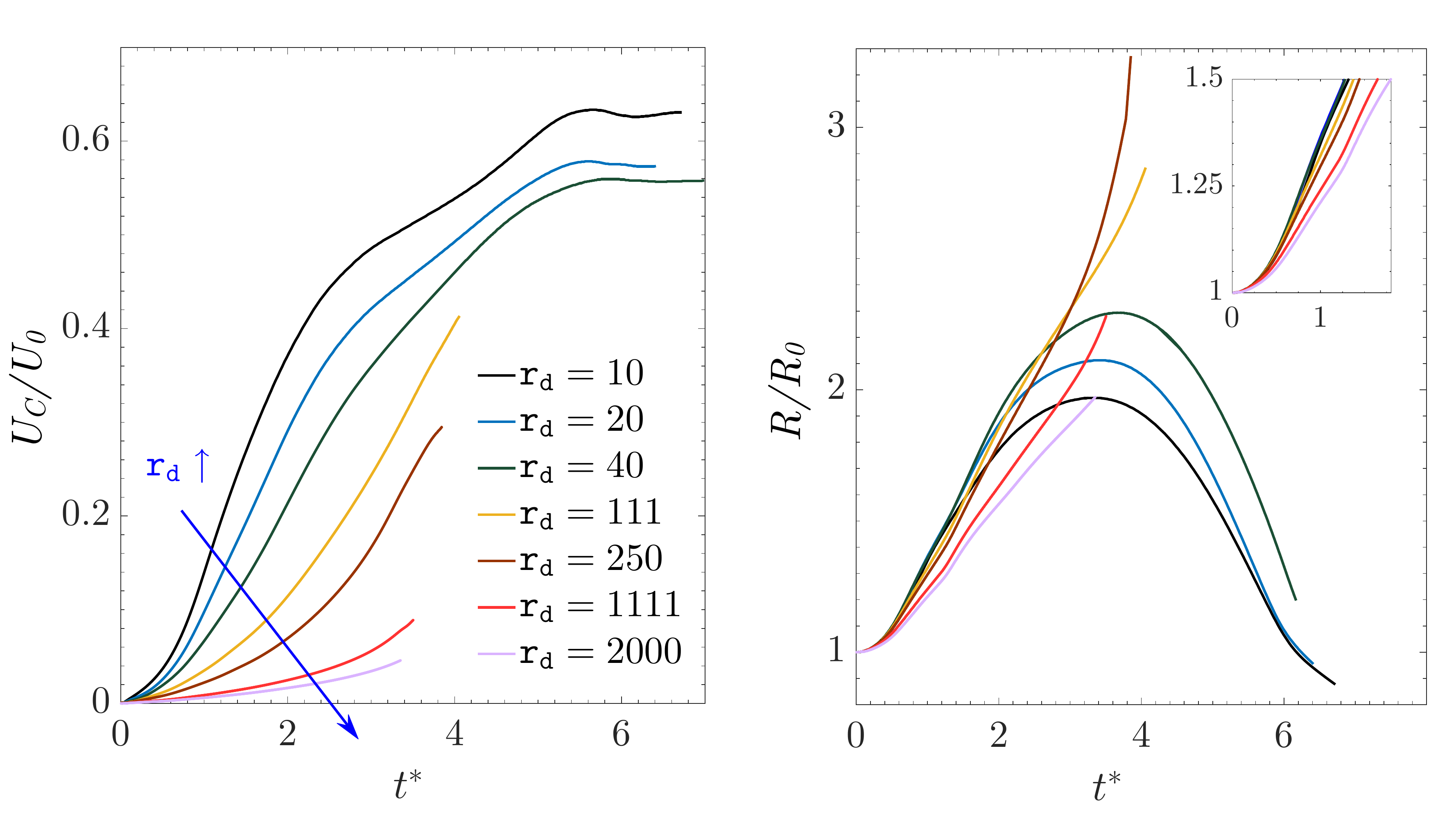}\\
\caption{\small{Centroid axial velocity \textit{(left)} and maximal drop radius \textit{(right)} as a function of the reduced time $t^*$, for density contrasts $\tt r_d$ in the $10-2000$ range and constant $\{\We=7.5; \,\Re=1090$; \, ${\tt r_v}=90.9\}$ (colours online).
}}
\label{fig:ux}
\includegraphics[width=0.8\textwidth]{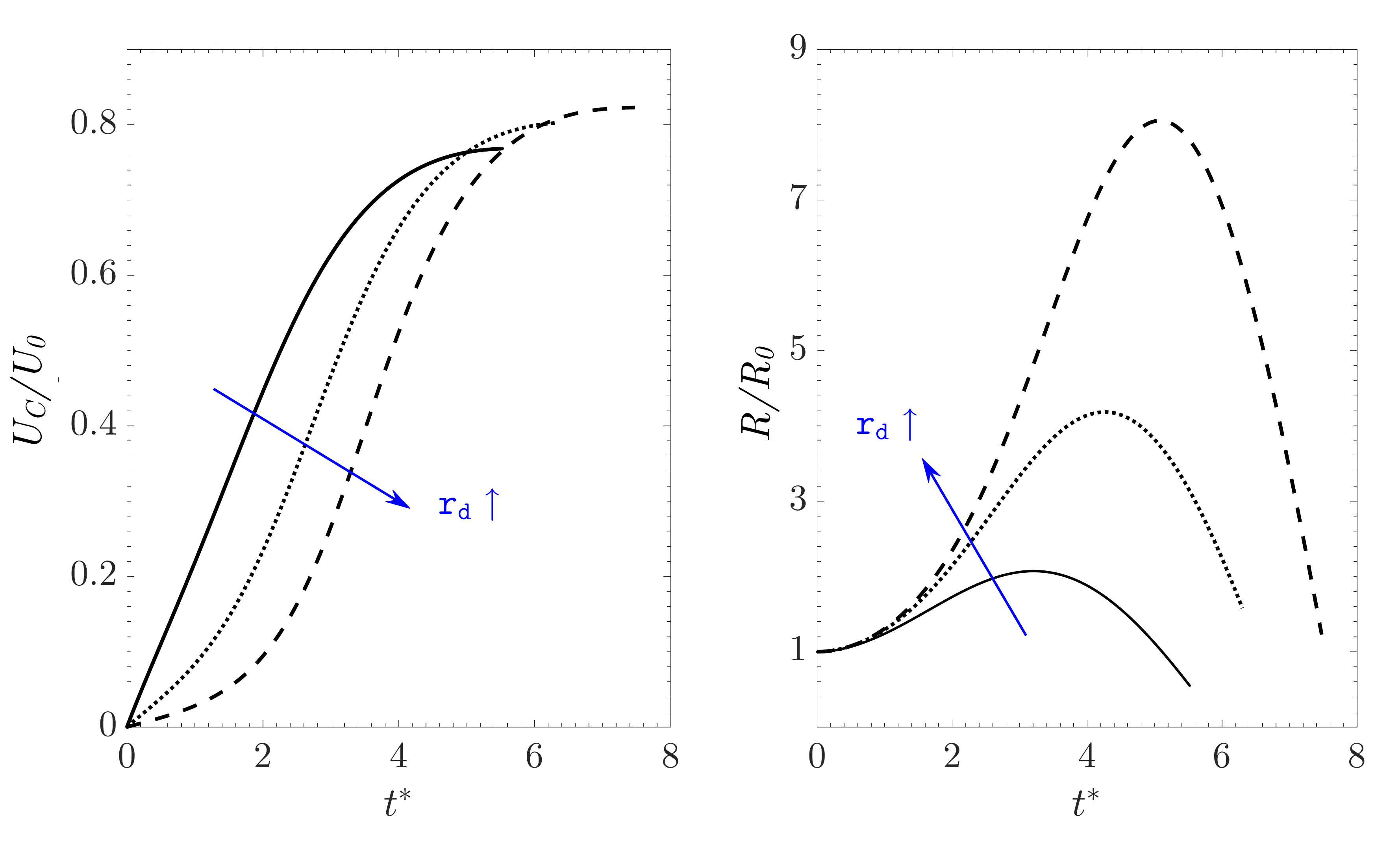}
\caption{\small{Centroid velocity \textit{(left)} and drop radius \textit{(right)} as a function of the reduced time $t^*$, as obtained by numerical integration of the simple, non-linear model (\ref{RR})-(\ref{ux}), for constant $\We=7.5$, and ${\tt r_d} = 10$ (solid line), ${\tt r_d} = 100$ (dotted line), ${\tt r_d} = 1000$ (dashed line). The proportionality coefficients in (\ref{RR})-(\ref{ux}) are arbitrarily set to $C_d=\alpha^2/4=1$.}}
\label{fig:modele}
\end{center}
\end{figure}

This dependency of the critical $\We$ number on the density ratio is not predicted by the existing linear models for the evolution of the drop radius \cite{Taylor63,VB09}, which for low Ohnesorge numbers predict fragmentation for $\We>{\We}_c$ regardless of the density contrast. Further, the exponential behaviour of the drop radius for $\We>{\We}_c$ theoretically predicted by such models hardly describes the early-time evolution of the drop maximal radius monitored in our simulations (as can be seen for example on figure \ref{fig:ux}\textit{(right)}).

However, a key element to understand the decrease in the fragmentation threshold from small to large density contrast (${\tt r_d}=10-250$) lies in the loss of the reduced time $t^*$-universality which we observe in the evolution of the (streamwise) centroid velocity $U_C$, shown in figure \ref{fig:ux}\textit{(left)} for constant $\{\We=7.5; \,\Re=1090$; \, ${\tt r_v}=90.9\}$, and for ${\tt r_d}$ varying in the $10-2000$ range. Whereas $t^*$ remains the relevant, typical timescale describing the acceleration of the centroid, figure \ref{fig:ux}(a) shows a non-ambiguous tendency for the drop centroid velocity to increase faster and {reach sooner a (temporary) saturation plateau} at smaller density contrasts. Interestingly, numerical results of \cite{meng01}, who used a compressible flow model at infinite $\Re$ and $\We$ to study drop breakup behind a shock-wave, are consistent with this observation despite important differences in flow conditions and modeling. Indeed their results for the evolution of the drop centroid velocity as a function of the reduced time $t^*$ show a monotonic increase of the early centroid acceleration for increasingly high Mach numbers (figure 14 in \cite{meng01}), \textit{i.e.} for increasingly high gas density behind the shockwave (following Rankine-Hugoniot jump conditions) and thus increasingly small density contrasts.

The corresponding evolutions of the drop radius $R$ are shown in figure  \ref{fig:ux}\textit{(right)}, with a blowup on early times in the figure inset: as in figure  \ref{fig:ux}\textit{(left)}, interrupted timeseries for ${\tt r_d} > 40$ denote (numerical) bursting. For all the density contrasts, and whether fragmentation eventually occurs or not, the evolution of the drop radius is similar up to $t^* \sim 1$: except for the two largest contrasts (${\tt r_d}= 1000-2000$) that yield slightly slower drop stretching, the different timeseries (almost) collapse onto a single curve once expressed in terms of the reduced time $t^*$. After this first stretching stage, the successive inflection points displayed by the various timeseries suggest that the drop radius initiates oscillations with growing period and amplitude as the density contrast increases from ${\tt r_d}=10$ up to (at least) ${\tt r_d}=250$, which in the cases where fragmentation occurs (here for ${\tt r_d} > 40$) are finally interrupted by a new growth and rapid bursting as in figure \ref{fig:B24}.

In fact, we can show that the smaller inertia of the lighter drops accounts both for the stabilising effect of small density contrasts and for the (non-exponential) time-evolution of the radius monitored in our simulations. To that effect, we assimilate the flattening drop with a disk of uniform thickness $e$, radius $R$ and constant volume ${\cal V}=4/3\pi R_0^3$, and use cylindrical coordinates to describe the fluid axisymmetric motion. Conservation of momentum in the streamwise direction can be approximated by
\be
\rho_L \frac{d}{dt} \left( U_C \cal{V} \right) = C_d \, \rho_G (U_0-U_C)^2  \pi R^2,
\label{dux}
\ee
where $C_d$ is the drag coefficient determined by the drop shape, so that the centroid acceleration evolves as the square of the drop radius (and vanishes as soon as the drop has reached the gas stream velocity). {Note that $C_d$ is in truth a time-dependent quantity and is expected to increase as the drop deforms (from a near to spherical shape to a near to cylindrical one), resulting in a faster increase in the centroid velocity; for simplicity we will however assume that $C_d$ is constant in the present model.}

Using now cylindrical coordinates to describe the fluid axisymmetric motion, conservation of mass inside the expanding disk yields:
\be
r \pd_t e + \pd_r (r \bar{u}_r)=0,
\ee
where $\bar{u}_r(r,t)=\int_{-e/2}^{e/2} u_r(r,z,t) \, dz$ is the radial velocity flux (per unit length), hence $\bar{u}_r= r \,e  \dot{R}/R$. Following the approach of \cite{VB09}, we now approximate the pressure difference in the gas between the high pressure region at the stagnation point on the drop pole ($r=0$) and the lower pressure region at the drop equator ($r=R$) by that of a hyperbolic stagnation flow with stretching rate $\gamma=\alpha (U_0-U_C) /R_0$, where $\alpha$ is a shape coefficient, yielding $p_G(R)-p_G(0)=-\rho_G\gamma^2 R^2/8$. Importantly, we will however assume here that the relative velocity $U_0-U_C$ remains \textit{time-dependent}. 

The pressure difference in the liquid is estimated from the pressure difference in the gas: at the pole $r=0$, the pressure jump across the interface vanishes due to the negligible curvature of the latter. At the equator however, {this} pressure jump can be approximated by $2\sigma /e \sim \tfrac32 \sigma R^2/R_0^3$ according to Laplace's law, so that the pressure difference in the liquid across the drop radius is finally
\be
p_L(R)-p_L(0) \ \sim \ -\tfrac{\alpha^2}{8} \rho_G \, (U_0-U_C)^2 R^2/R_0^2 \ + \ \tfrac32 \sigma R^2/R_0^3.
\label{Dp}
\ee
Discarding viscous damping, the axisymmetric Euler equation for the radial velocity $u_r$ inside the drop writes
\be
\rho_L ( \pd_t u_r + u \cdot \nabla u_r) = - \pd_r p.
\label{euler}
\ee
Assuming uniform radial motion through the drop thickness (hence $u_r=\bar{u}_r/e= r \dot{R}/R$) and averaging (\ref{euler}) over the drop domain using (\ref{Dp}) yields
\be
 \frac{\ddot{R}}{R}  \  \sim  \, {\tt r_d}^{-1} \left( \frac{\alpha^2}{4} \, \Big(1-\fb{U_C}{U_0}\Big)^2 - \fb{3} \We\right) \, \frac{U_0^2}{R_0^2},
 \label{RR}
 \ee
an evolution equation for the drop radius non-linearly coupled with that of the centroid velocity derived from (\ref{dux}):
\be
\dot{U}_C \ \sim \ \fb{3}{4} C_d \, {\tt r_d}^{-1} \Big(1-\fb{U_C}{U_0}\Big)^2 \, \fb{U_0^2}{R_0^3} \, R^2,
\label{ux}
\ee
with $R(0)=R_0$ and $U_C(0)=0$. 
The non-linear coupling between (\ref{RR}) and (\ref{ux}) accounts for the role played by the density contrast in the fragmentation threshold. If we ignore the drop acceleration as in \cite{VB09}, as is realistic for large $\tt r_d$ at early times, (\ref{RR}) becomes linear in $R$. The drop radius then undergoes exponential stretching as soon as the Weber number exceeds the critical value ${\We}_c = 12/\alpha^2$, on a typical timescale determined by the reduced time {$t^*={\tt r_d}^{-1/2}\tilde{t}$}. \textcolor{black}{(In \cite{VB09} the shape coefficient determining the stretching rate of the hyperbolic stagnation flow in the gas is $\alpha=2$, an intermediate value between the case of a spherical obstacle ($\alpha=3$) and that of a disc ($\alpha=\pi/2$). This value of $\alpha$ yields ${\We}_c = 3$, in excellent agreement with the fragmentation threshold found numerically at large density contrast.)}
However, taking (\ref{ux}) into account modifies the critical Weber number: it is clear from (\ref{RR}) that an increase in $U_C$ achieved sufficiently fast over the reduced timescale can significantly raise the fragmentation threshold. The acceleration at early times ($\tilde{t}=0^+$) is larger for smaller density contrasts, so that a faster decrease in the right-hand side of (\ref{RR}) is expected as $\tt r_d$ decreases, associated with a smaller stretching amplitude and oscillation period (expressed in reduced timescale). 

Solutions of (\ref{RR})-(\ref{ux}) determined by numerical integration are shown in figure \ref{fig:modele} for constant $\We=7.5$ and three different density contrasts in the $10-1000$ range. Even though our simple non-linear model does not allow for a precise quantitative description of the drop deformation dynamics {and does not account for the slight increase in $\We_c$ numerically observed at ${\tt r_d}=1000-2000$ (presumably because of the important simplifications made in the derivation of (\ref{RR})-(\ref{ux}))}, the principal features of figure \ref{fig:ux} - stronger acceleration and earlier {(temporary)} saturation of the drop velocity for small density contrasts, as well as shorter and weaker oscillations of the drop radius -  are well reproduced. Our findings thus demonstrate that the variation in the drop inertia indeed accounts for the dependency of the deformation dynamics on the density contrast over the reduced timescale $t^*$, and for the stabilising effect of small density contrasts with respect to fragmentation. However, it is important to emphasize that our model does not address the deformation and bursting dynamics during the later inflation stage. \textcolor{black}{Moreover, our axisymmetric simulations cannot address the inherently three-dimensional mechanisms of sheet piercing in the inflating drop, which would require additional (and computationally much more demanding) studies.} Full understanding of the bursting mechanisms, including the possible development of hydrodynamic instabilities in the rim or and the thinning film, is beyond the scope of the present article and remains a challenging problem for future investigation.


\section{Conclusion}

Our axisymmetric simulations of impulsively-accelerated drop breakup show for the first time a significant influence of the liquid-gas density contrast on the fragmentation thresholds at low Ohnesorge number. The density contrast was varied over two decades (${\tt r_d}=10-2000$), showing good agreement with previous numerical and experimental works. We suggest a simple theoretical argument to determine the critical Weber number characterising the transition from bursting to stripping regime, which we find to increase for decreasing density contrasts, consistently with our numerical results. {Although the objective of the present paper was not to provide a review of the studies that could experimentally assess the validity of these theories over the whole parameter space (large and small Reynolds and Weber numbers, and large and small density/viscosity contrasts), we believe that the full range of such experimental investigations remains to be done and would be of great interest.} Finally, we explain the significant rise of the fragmentation threshold observed here at small density contrasts by the nonlinear coupling between the stretching drop radius and the {centroid acceleration}. Our results demonstrate the need for adaptation of existing breakup criterion, \textcolor{black}{such as those currently used in non-DNS CFD models,} so as to take into account the effect of density contrast on the fragmentation threshold and regimes.

\vspace{5ex}

{\it Acknowledgement.} The authors wish to thank Sasha Korobkin, St\'ephane Popinet, Wojciech Aniszewski and Christophe Josserand for fruitful discussions.

\bibliographystyle{plain}
\bibliography{fragmentation.bib}

\end{document}